\documentclass[aps,prd,preprintnumbers,reprint,nofootinbib,showpacs,onecolumn,notitlepage]{revtex4-1}
\usepackage{graphicx,bm,amssymb,amsmath}
\usepackage[normalem]{ulem}
\usepackage{color}

\newcommand{\defeq}{\mathrel{\mathop:}=}

\usepackage[mathletters]{ucs}
\usepackage[utf8x]{inputenc}

\begin{document}

\title{\boldmath Addendum: Electroweak Sudakov effects in $W$, $Z$ and $\gamma$ production at large
transverse momentum}
\author{Thomas Becher}
\affiliation{Albert Einstein Center for Fundamental Physics, Institut f\"ur Theoretische Physik, Universit\"at Bern,
  Sidlerstrasse 5, CH-3012 Bern, Switzerland}
\author{Xavier \surname{Garcia i Tormo}\\ \phantom{a}}
\affiliation{Albert Einstein Center for Fundamental Physics, Institut f\"ur Theoretische Physik, Universit\"at Bern,
  Sidlerstrasse 5, CH-3012 Bern, Switzerland}

\date{\today}


\begin{abstract}
In this addendum to Ref.~\cite{Becher:2013zua}, we give results for the electroweak Sudakov corrections in gauge-boson production at large transverse momentum $p_T$ at proton colliders. In order for the results to be easily usable, we provide a simple and accurate parameterization of the corrections as a function of $p_T$ and the center-of-mass energy $\sqrt{s}$. Additionally, we also discuss the dependence of the electroweak corrections on the rapidity of the produced boson, and comment on the complications that arise in the photon-production case due to isolation requirements. \end{abstract}


\maketitle

In Ref.~\cite{Becher:2013zua} we presented a study of electroweak Sudakov corrections to the cross sections for single $W$, $Z$, and $γ$ production at large transverse momentum $p_T$, within the framework of Soft Collinear Effective Theory (SCET). Our results are based on a factorization formula for the hadronic cross section near the partonic threshold and include both Quantum Chromodynamics (QCD) and electroweak corrections. To quantify the effect of including electroweak corrections in the cross section, it is convenient to look at the relative difference $Δσ^{ew}$ of the cross section computed with and without electroweak effects, i.e. 
\begin{equation}\label{eq:diffsigew}
\Delta\sigma^{ew}\defeq\frac{\sigma_{ew}^{i}-\sigma^{i}}{\sigma^i},
\end{equation}
where $σ^i$ represents the cross section computed including only QCD effects, at a certain order denoted by $i$, and $σ_{ew}^i$ represents the same cross section including electroweak effects. As shown in Ref.~\cite{Becher:2013zua}, the relative importance of the electroweak corrections, given by $Δσ^{ew}$, is essentially independent of the order $i$ which we use to compute the QCD effects. This implies that, to good accuracy, one can include electroweak effects in existing pure-QCD computations simply as an overall multiplicative prefactor $1+Δσ^{ew}$. 

The inclusion of electroweak Sudakov effects is essential in order to obtain precise predictions for the $p_T$ spectrum in the region $p_T\gg M_{W,Z}$, since their effect can be as large as $20\%$ for $p_T\sim 1~\rm{TeV}$. Therefore, any LHC analysis of these processes, as well as studies of future colliders at higher center-of-mass energies, must include them. In Ref.~\cite{Becher:2013zua} we presented the correction factor $Δσ^{ew}$ for a center-of-mass energy $\sqrt{s}=7~\rm{TeV}$. Results at $\sqrt{s}=33~\rm{TeV}$ and $\sqrt{s}=100~\rm{TeV}$ were presented in Ref.~\cite{Mishra:2013una}, in the context of the Snowmass study of future colliders. It was found that the relative correction $Δσ^{ew}$ depends only mildly on the center-of-mass energy. For further Run I and future Run II LHC analyses it would be useful, to have an expression that allows one to obtain $Δσ^{ew}$ for different values of $\sqrt{s}$. Providing such an expression is the main motivation for this addendum. In order to obtain it, we compared the correction factors $Δσ^{ew}$ at different $\sqrt{s}$ values and found that their ratios are largely independent of $p_T$. For this reason, we first provide a simple parameterization of the results at $\sqrt{s}=7~\rm{TeV}$, which serves as a reference, and then supply the necessary prefactor to convert them to other center-of-mass energies.  

\begin{figure}
\includegraphics[width=10cm]{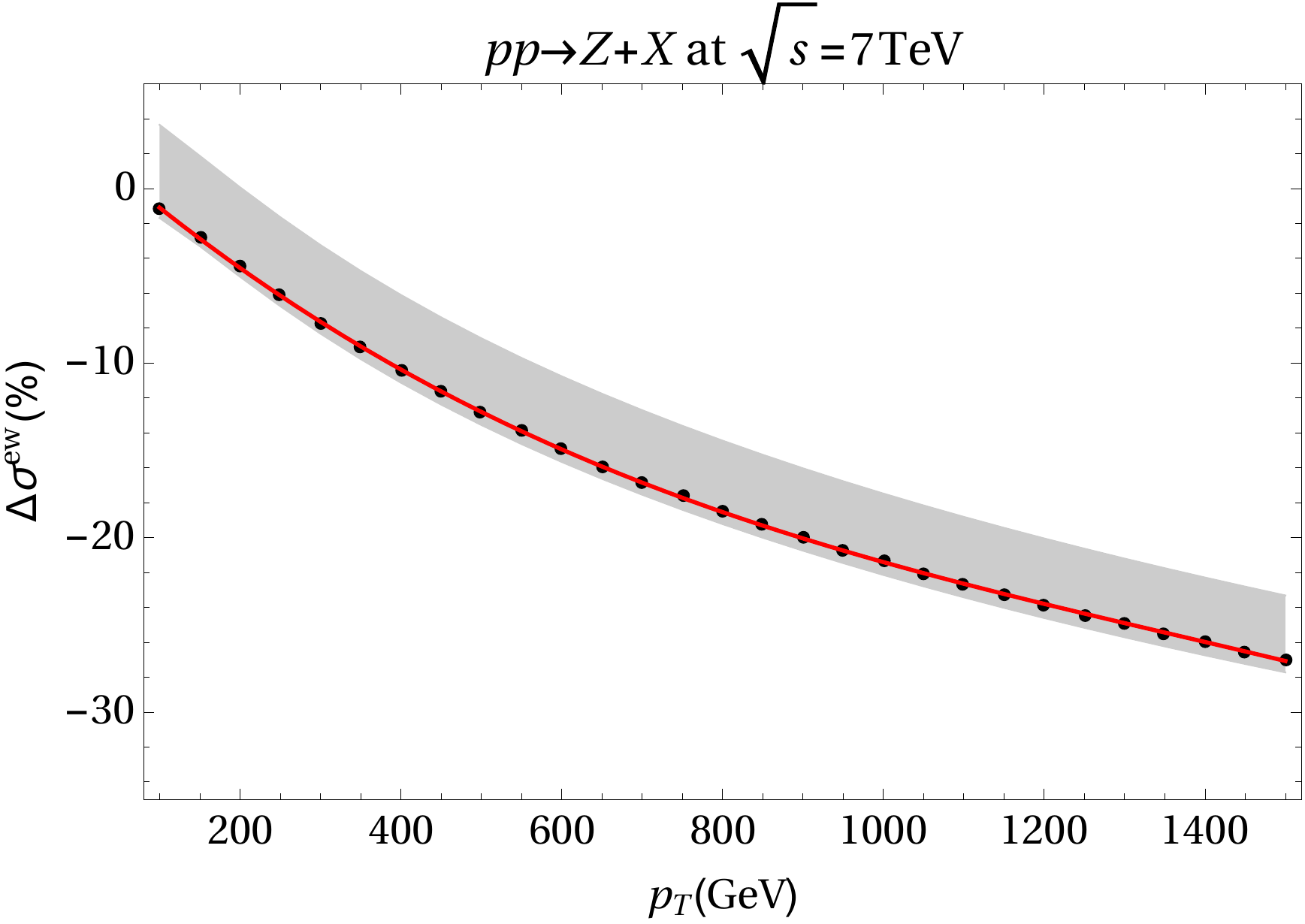}
\caption{$Δσ^{ew}$ for $Z$ production at $\sqrt{s}=7~\rm{TeV}$ computed using the exact expressions from Ref.~\cite{Becher:2013zua}, black points, and the parameterization in Eq.~(\ref{eq:polpt}), red line. For reference, we also show the theoretical uncertainty of the result as the grey band (see Ref.~\cite{Becher:2013zua} for further details).} 
\label{fig:Z7TeV}
\end{figure}

The dependence on $p_T$ of $Δσ^{ew}$ at our reference scale $\sqrt{s_0}=7~\rm{TeV}$ can be very accurately parameterized by a third-order polynomial, according to
\begin{equation}\label{eq:polpt}
Δσ^{ew}_V(p_T,s_0)=a_0^V+a_1^Vx+a_2^Vx^2+a_3^Vx^3=:g_V(p_T),
\end{equation}
where $x$ represents the $p_T$ of boson $V$ expressed in TeV, and the $a_i^V$ coefficients for the different gauge bosons are given in Tab.~\ref{tab:ptdep}, both for the central value and the associated error band. The accuracy of the parameterization in Eq.~(\ref{eq:polpt}) is better than 1$\%$ for most of the $p_T$ values between 100 and 1500~GeV, and the difference between the parameterization and the exact result is always much smaller than the theoretical uncertainties of $Δσ^{ew}$. Eq.~(\ref{eq:polpt}) can therefore be safely used in the whole $p_T$ range we study, but should not be employed beyond $p_T=1500$~GeV. To illustrate this, we show in Fig.~\ref{fig:Z7TeV} the exact result for $Δσ^{ew}$ (black points) together with the outcome of Eq.~(\ref{eq:polpt}) (red line), as well as the theoretical uncertainty due to scale variations (grey band). 

\begin{table*}[b]
\begin{tabular}{ccccc}
\begin{tabular}{|c||c|c|c|c|}\hline
& \multicolumn{4}{|c|}{central value}  \\ \hline
& $a_0^V$ & $a_1^V$ & $a_2^V$ & $a_3^V$ \\ 
\hline
$Z$ & 2.763     & -40.76     & 21.95     & -5.356    \\
\hline
$γ$ & 1.713 & -21.68 & 12.16 & -3.050 \\
\hline
$W^+$ & 3.816  & -45.25 & 23.74 & -5.899 \\
\hline
$W^-$ & 4.074 & -47.60  & 25.97 & -6.414 \\
\hline
\end{tabular}
&  \hspace{0.8cm}  &
\begin{tabular}{|c||c|c|c|c|}\hline
& \multicolumn{4}{|c|}{upper edge}  \\ \hline
& $a_0^V$ & $a_1^V$ & $a_2^V$ & $a_3^V$ \\ 
\hline
$Z$ & 7.640     & -42.41     & 22.88     & -5.609    \\
\hline
$γ$ & 4.556 & -22.06 & 12.30 & -3.072 \\
\hline
$W^+$ & 9.592  & -46.81 & 24.09 & -5.976 \\
\hline
$W^-$ & 9.906 & -49.00  & 26.63 & -6.590 \\
\hline
\end{tabular}
& \hspace{0.8cm} &
\begin{tabular}{|c||c|c|c|c|}\hline
& \multicolumn{4}{|c|}{lower edge}  \\ \hline
& $a_0^V$ & $a_1^V$ & $a_2^V$ & $a_3^V$ \\ 
\hline
$Z$ & 2.302    & -41.49     & 22.53     & -5.504    \\
\hline
$γ$ & 0.824 & -23.23 & 13.44 & -3.428 \\
\hline
$W^+$ & 2.307  & -45.99 & 24.97 & -6.244 \\
\hline
$W^-$ & 2.622 & -48.51  & 26.84 & -6.637 \\\hline
\end{tabular}
\end{tabular}

\caption{Values of the $a_i^V$ coefficients in Eq.~(\ref{eq:polpt}), parameterizing the $p_T$ dependence of $Δσ^{ew}_V$ at $\sqrt{s}=7~\rm{TeV}$, for the different gauge bosons $V$.}
\label{tab:ptdep}
\end{table*}

\begin{figure}
\includegraphics[width=8cm]{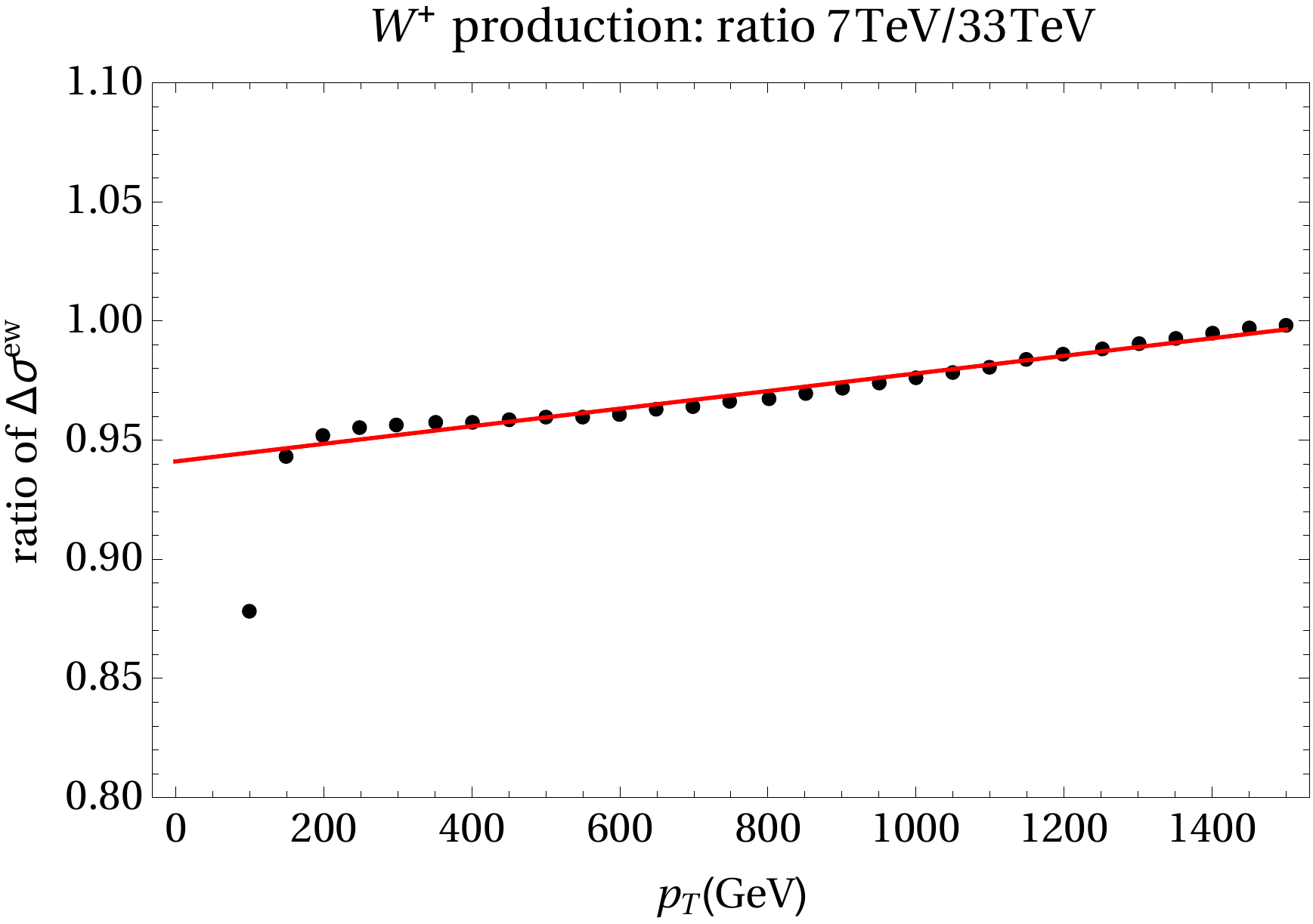}
\includegraphics[width=8cm]{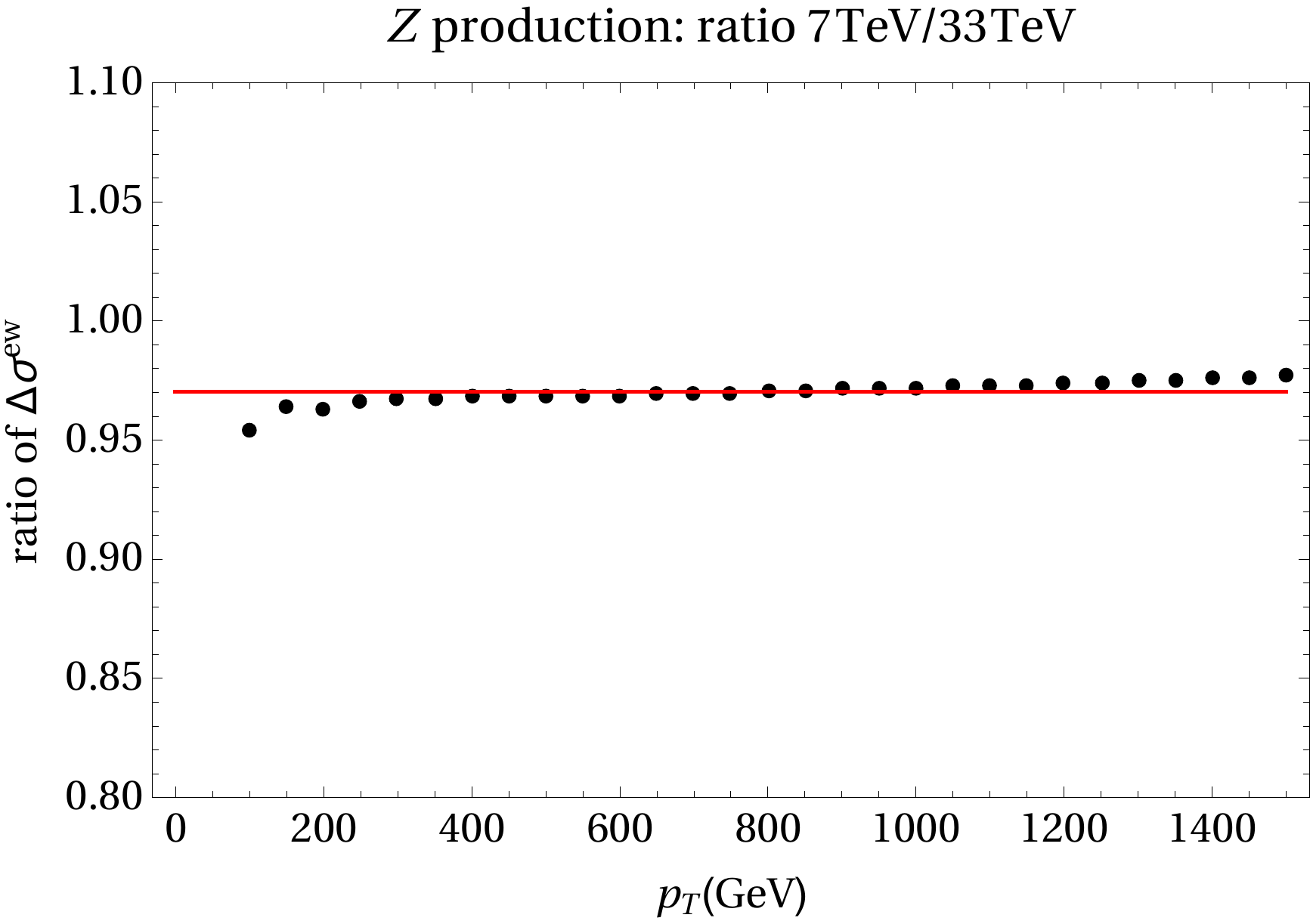}
\caption{Ratios of $Δσ^{ew}$ at different energies, for a $pp$ collider. Left: $\sqrt{s}=7\,\rm{TeV}$ over $\sqrt{s}=33\,\rm{TeV}$ for $W^+$. Right: $\sqrt{s}=7\,\rm{TeV}$ over $\sqrt{s}=33\,\rm{TeV}$ for $Z$. The black points are always the exact result using the expressions from Ref.~\cite{Becher:2013zua}, and the red lines are the $f_V$ factors in Eqs.~(\ref{eq:fZs})-(\ref{eq:fWps}).} 
\label{fig:ratios}
\end{figure}

Next, we provide prefactors $f_V(\sqrt{s})$ that convert Eq.~(\ref{eq:polpt}) to other center of mass energies, according to
\begin{equation}
Δσ^{ew}_V(p_T,s)=g_V(p_T)/f_V(\sqrt{s}) \,,
\end{equation}
where $Δσ^{ew}_V(p_T,s)$ represents the difference $Δσ^{ew}_V$ at a center-of-mass energy $\sqrt{s}$, expressed in TeV. The $f_V(\sqrt{s})$, for each of the different bosons, are given by:
\begin{eqnarray}
f_Z(\sqrt{s}) & = & 1-1.0376\times10^{-2}y+5.4060\times10^{-4}y^2,\label{eq:fZs}\\
f_γ(\sqrt{s}) & = & 1-2.3355\times10^{-2}y+1.2310\times10^{-3}y^2,\\
f_{W^-}(\sqrt{s}) & = & 1-9.5067\times10^{-3}y+4.9809\times10^{-4}y^2,\\
f_{W^+}(\sqrt{s}) & = & 1-1.9924\times10^{-2}y+1.0891\times10^{-3}y^2\nonumber\\
&&+p_T\left(1.2802\times10^{-2}y-7.6955\times10^{-4}y^2\right),\label{eq:fWps}
\end{eqnarray}  
where $y=(\sqrt{s}-\sqrt{s_0})/\sqrt{s_0}$. Only in the case of $W^+$ one observes a slight dependence of the ratio on $p_T$ and for this reason we decided to include also a term linear in $p_T$ in $f_{W^+}$; it is understood that $p_T$ is expressed in TeV. Using the $f_V$ factors to obtain $Δσ^{ew}$ at different energies provides an accuracy better than 1$\%$ for most of the $p_T$ values between 100 and 1500~GeV, and the difference with the exact result is always much smaller than the theoretical uncertainties. We illustrate this in Fig.~\ref{fig:ratios}, by comparing ratios of $Δσ^{ew}$ computed at different energies using the exact result (black points) with the $f_V$ factors in Eqs.~(\ref{eq:fZs})-(\ref{eq:fWps}) (red line). One can also use the $f_V$ factors above to convert the upper and lower limits of the uncertainty bands at 7~TeV to other $\sqrt{s}$ values.

Our finding that $Δσ^{ew}$ given above has only a mild dependence on $\sqrt{s}$, is a reflection of the fact that most of the electroweak corrections arise in the hard function in the factorization formula, as discussed in Ref.~\cite{Becher:2013zua}. The hard function evolves from its characteristic scale, which is of order $p_T$, to the factorization scale, which is of order $M_Z$, and the electroweak corrections generated by this evolution are independent of $\sqrt{s}$. We  note that the predictions of our effective-theory setup become less accurate as $p_T\to M_{Z,W}$, since the hierarchy of scales on which it is based is no longer present. In particular we cannot use it to compute electroweak corrections below $p_T=M_{Z,W}$ and, consequently, one should just set our $Δσ^{ew}$ factor to 0 for $p_T\lesssim M_{Z,W}$. Note also that we estimated the uncertainties of our results by varying the different scales appearing in the factorization formula independently. When $p_T\to M_{Z,W}$ the different scales are not really independent and this procedure may result in an overestimate of the errors, but since the method itself deteriorates in this limit we recommend to use the error bands given above all the way down to $p_T=M_{Z,W}$.

\begin{figure}
\includegraphics[height=5.5cm]{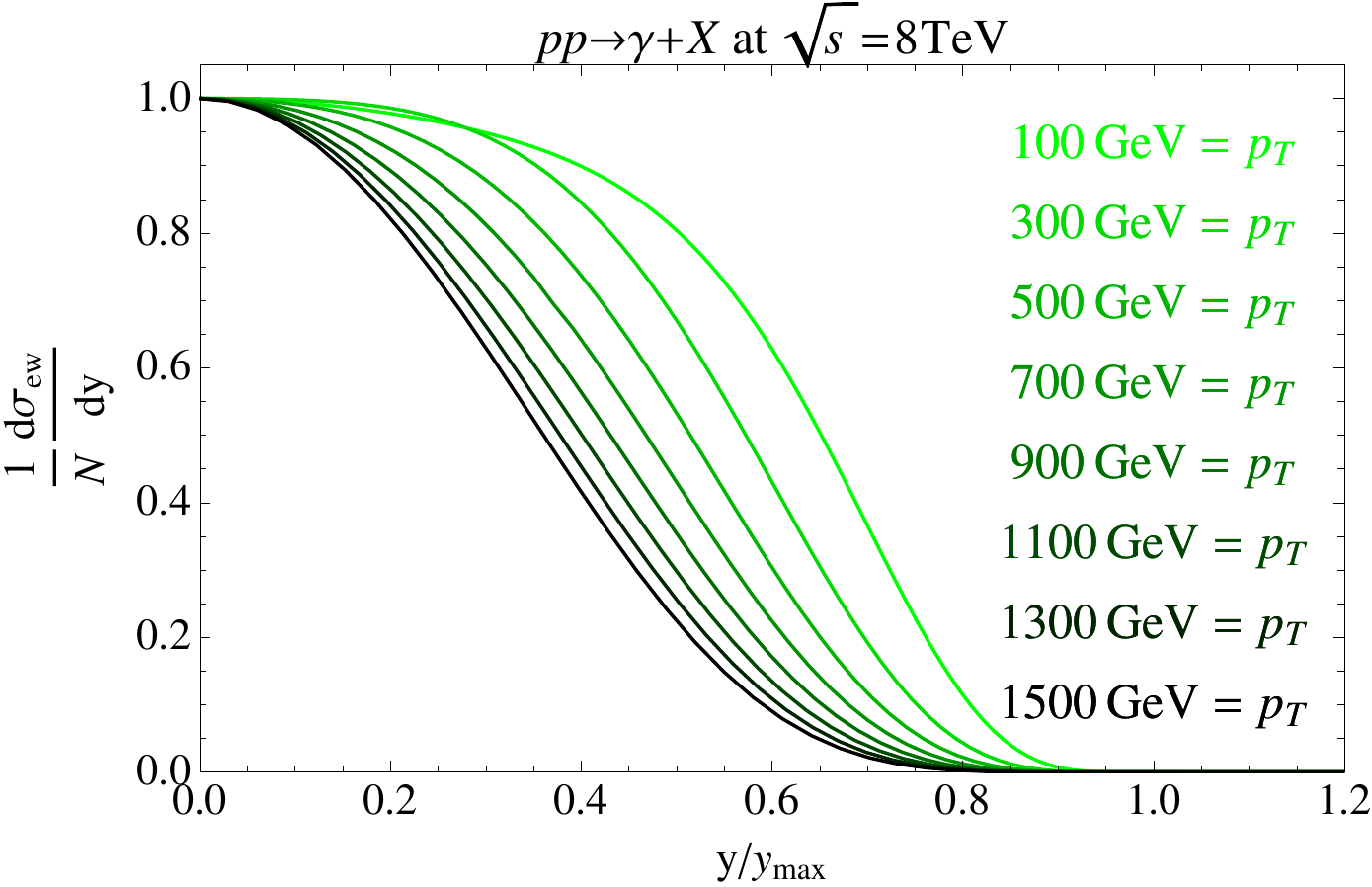}\hspace{0.1cm}
\includegraphics[height=5.5cm]{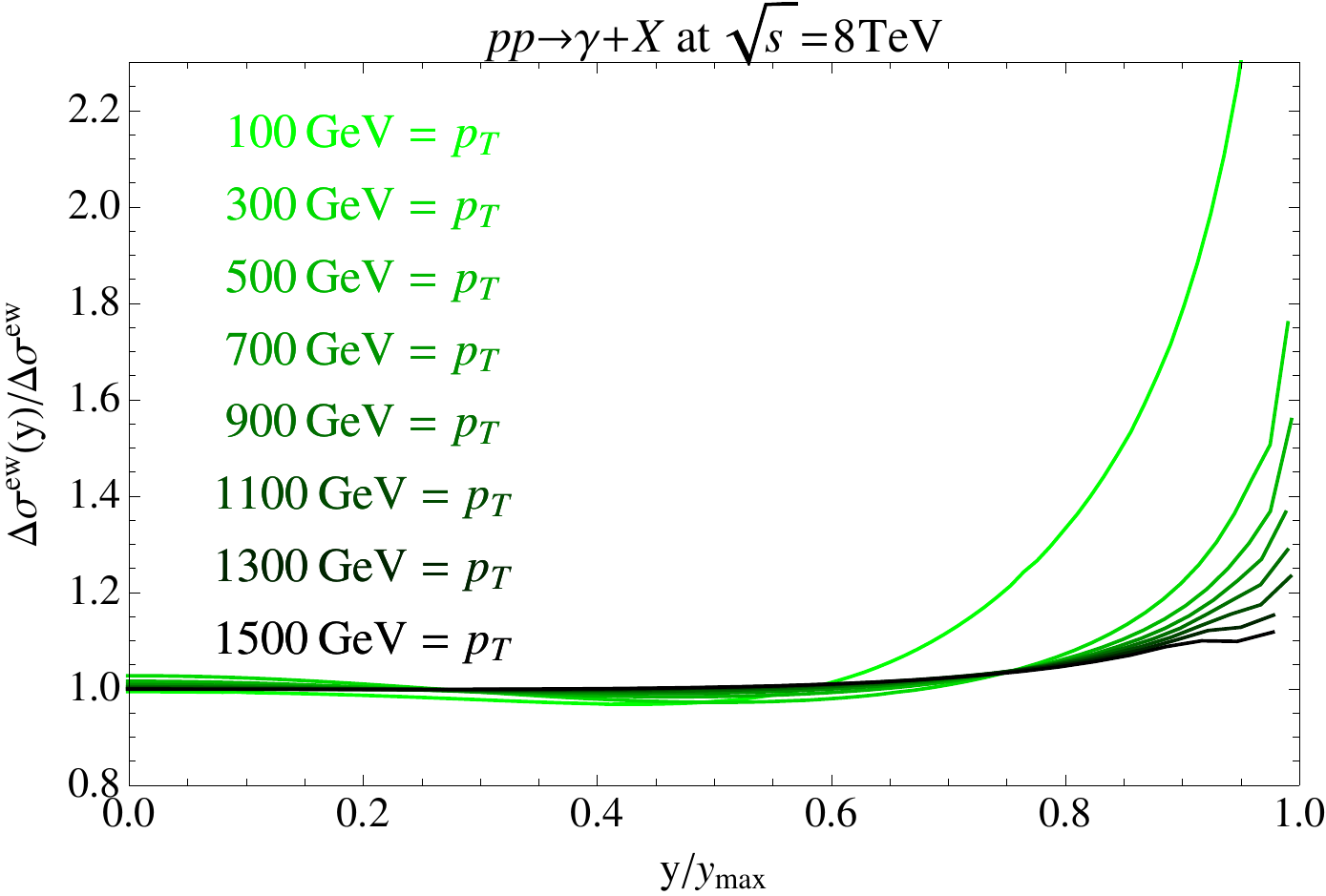}
\caption{Left panel: Cross section differential in rapidity, normalized to the value at $y=0$. Right panel: Size of the electroweak corrections as a function of rapidity, normalized to the value of the correction integrated over rapidity.}
\label{fig:rapdep}
\end{figure}

\begin{figure}[t]
\includegraphics[width=7.2cm]{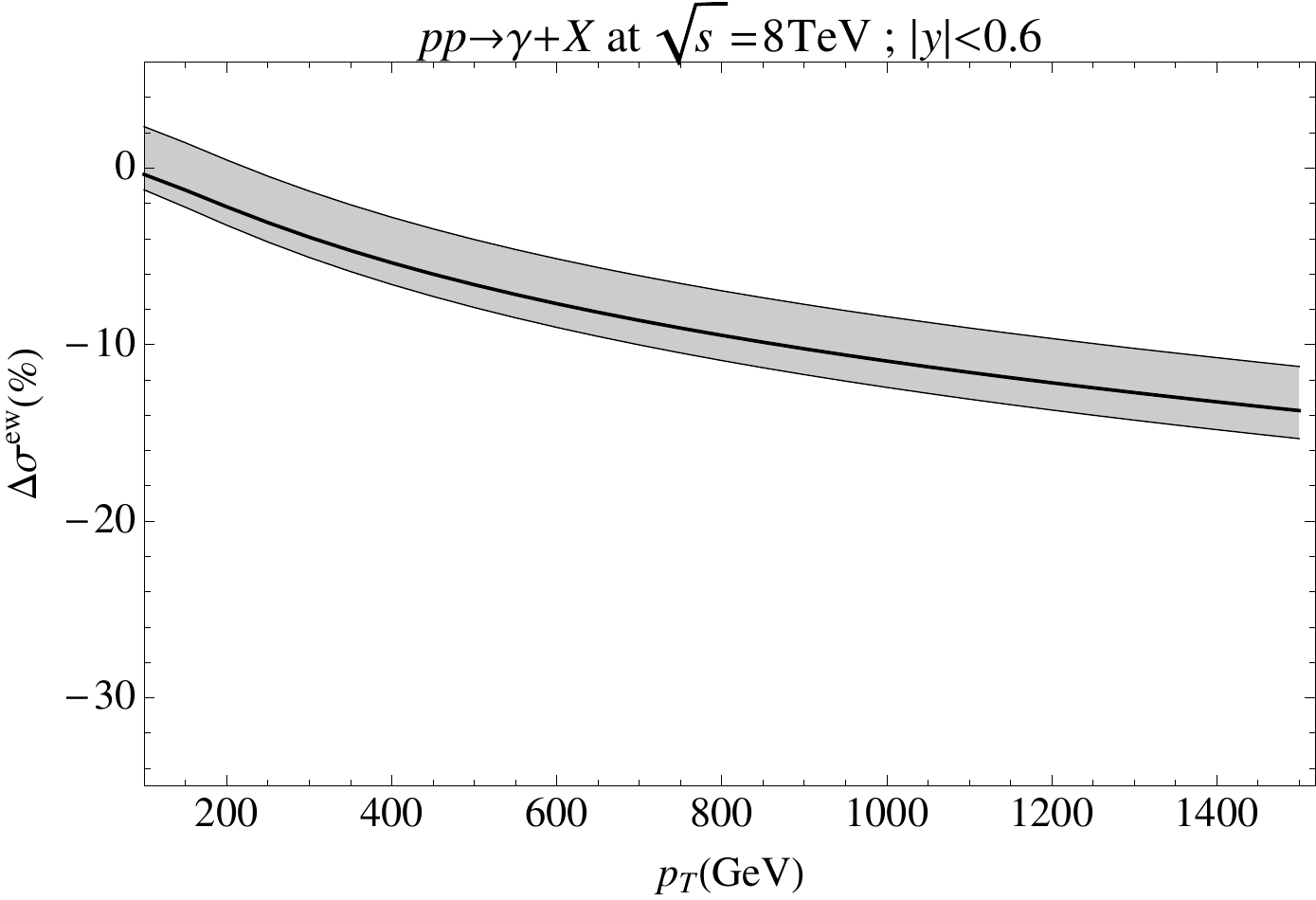} \hspace{0.5cm}
\includegraphics[width=7.2cm]{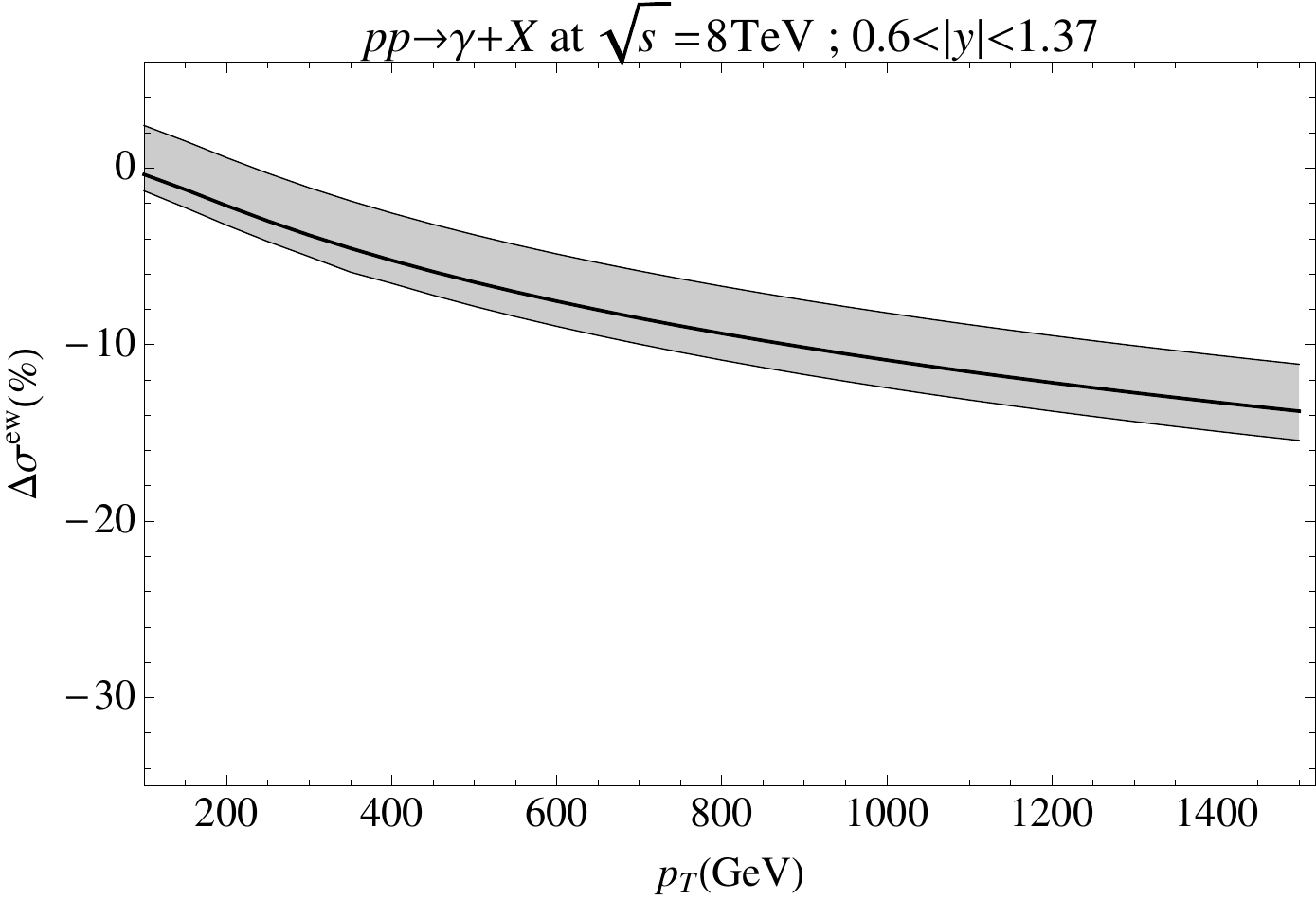}
\includegraphics[width=7.2cm]{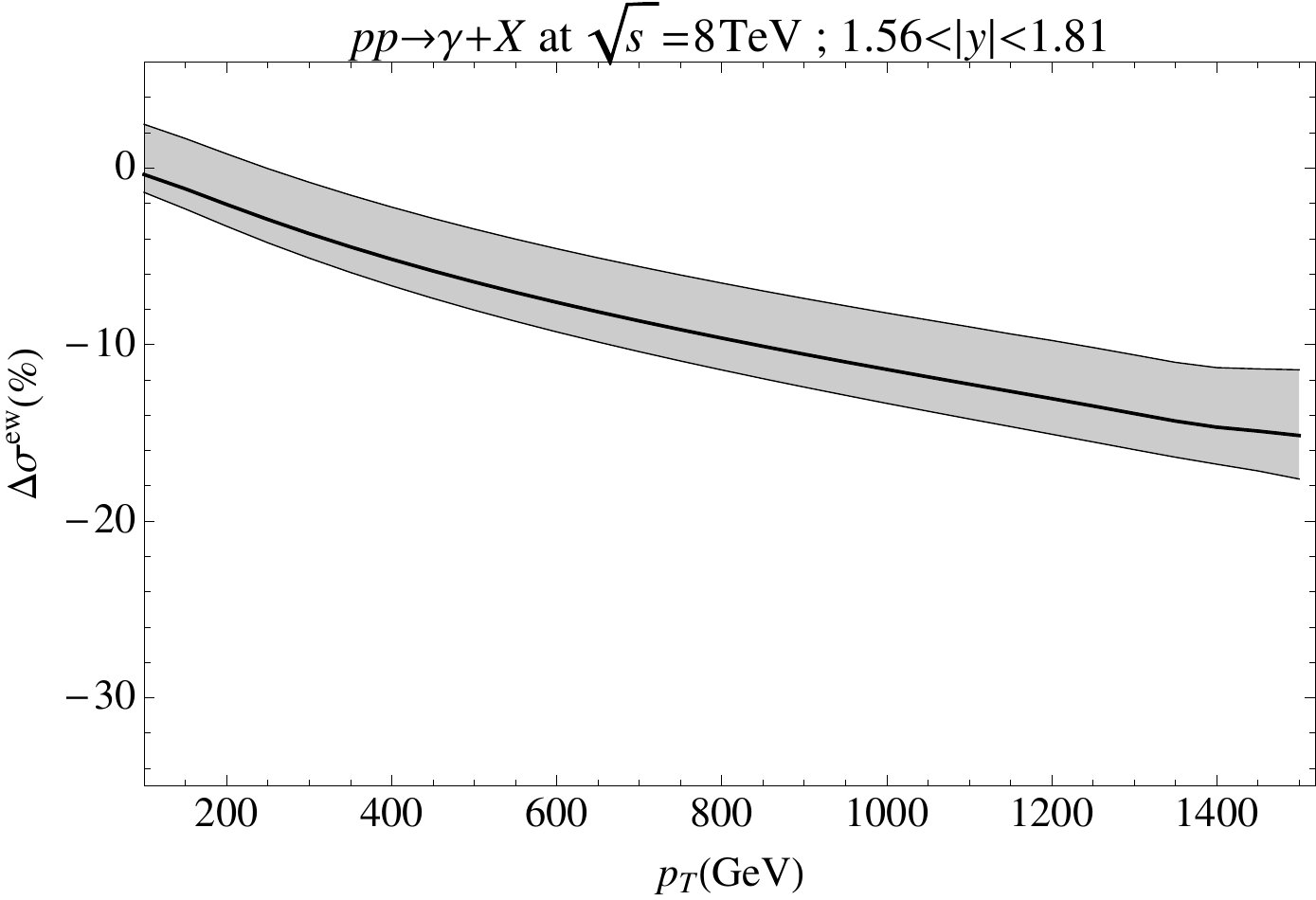} \hspace{0.5cm}
\includegraphics[width=7.2cm]{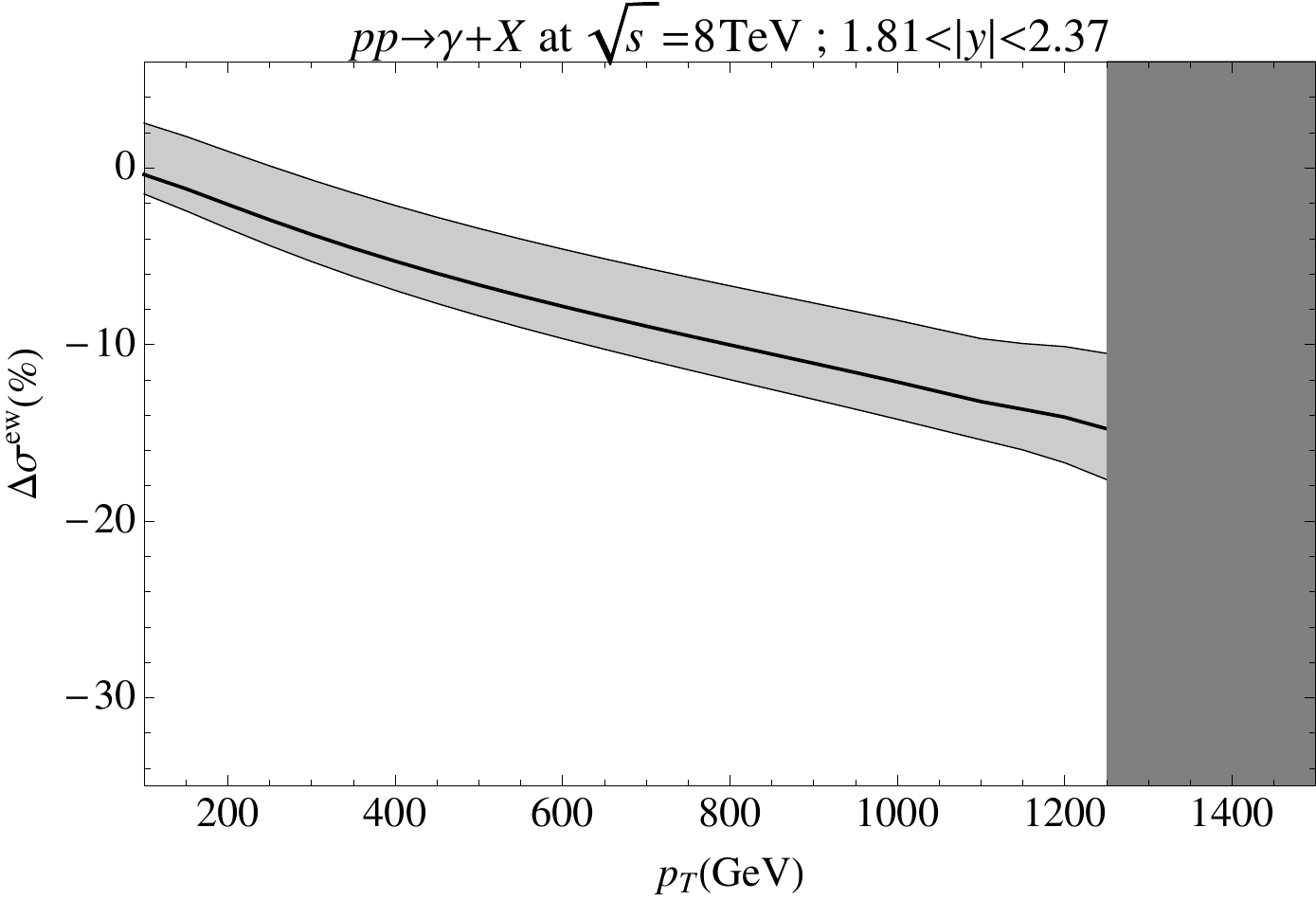}
\caption{$Δσ^{ew}$ at 8~TeV for γ production at different rapidity ranges. In the highest rapidity range shown in the last plot the cross section is essentially 0 for $p_T>1200$~GeV.}
\label{fig:rapranges}
\end{figure}

Our approximate results are for cross sections integrated over the rapdity of the bosons, but to good accuracy the corrections can also be used for cross sections with rapidity cuts. To understand this, consider Fig.~\ref{fig:rapdep}, which shows the dependence of the corrections on rapidity for a few fixed $p_T$ values. The left panel of the figure shows the cross section differential in rapidity, normalized to $N=dσ/dy|_{y=0}$, the value of the cross section at $y=0$. The right panel shows the size of the electroweak corrections $Δσ^{ew}(y)$ as a function of rapidity, normalized to $Δσ^{ew}$ integrated over rapidity. In the horizontal axis, the rapidity is normalized to the maximum value $y_{\rm max}$, which is given by
\begin{equation}
y_{\rm max}=-\ln\frac{s+M_V^2-\sqrt{(M_V^2-s)^2-4p_T^2s}}{2\sqrt{s}\sqrt{p_T^2+M_V^2}} \approx \frac{1}{2}\ln\frac{s}{p_T^2}\,,
\end{equation}
where $M_V$ is the mass of the produced boson.
Only positive rapidities are shown because the cross section is symmetric in rapidity for $pp$ collisions. From the figure we learn that, first of all, the bulk of the cross section comes from the central rapidity region, $y/y_{\rm max}\lesssim0.8$, and, secondly, that the electroweak corrections are essentially flat in this region. Therefore, $Δσ^{ew}$ is mostly independent of rapidity. Furthermore, the deviations from unity in Figure~\ref{fig:rapdep} mostly arise for bosons at low $p_T$ and therefore correspond to very high rapidities (e.g. at $p_T=300\, {\rm GeV}$ and $\sqrt{s} = 8 \,{\rm TeV}$, $y_{\rm max} \approx 3.3$) which are typically not included in the experimental measurements. The fact that the corrections are largely independent of the vector-boson rapidity is demonstrated in Fig.~\ref{fig:rapranges}, where we have computed them for different rapidity ranges relevant in the ATLAS experiment and find that they are almost identical in every case.

As a final, important point, we briefly want to discuss the effect of experimental cuts beyond the boson rapidity. Our computation is based on threshold resummation, i.e. it assumes that the final state only contains the hard vector boson, recoiling against a low-mass jet, plus soft gluons and soft photons. In experimental measurements a variety of phase-space cuts are imposed and while we hope to capture the bulk of the EW corrections, it is clear that the setup of our calculation is not fully realistic. In particular, the EW corrections can change significantly if the experimental setup is such that additional, unobserved massive EW bosons can be radiated into the final state since their contributions can cancel some of the EW Sudakov logarithms. A second issue related to experimental cuts is that near the partonic threshold, the fragmentation contribution in photon production is power suppressed \cite{Becher:2009th}, but in practice one needs to define what is meant by a final-state photon and depending on the photon isolation requirements, a fragmentation contribution will also arise and will involve different EW corrections. The experimental definition of the photon final state will also affect which value of electromagnetic coupling should be used when computing the the cross section. (The value is obviously irrelevant for the ratio in Eq.~\eqref{eq:diffsigew} we present in this paper, but its choice is important when computing the cross section.) In our computation, the appropriate value of the electromagnetic coupling is the value at the invariant mass of the vector-boson final state. If the experiment would be able to resolve photon jets with arbitrary small mass, the proper value would be $\mu=0$ and one would use the on-shell, low energy value of $\alpha_{e.m.}\approx 1/137$, see e.g. the discussion in \cite{Czarnecki:1998tn}. However, with the photon isolation requirements at the LHC, the photon-jet has a non-zero mass and a value such as $\alpha_{e.m.}(M_Z)\approx 1/128$ might be more appropriate.

{\em Acknowledgements:} 
We thank Marat Freytsis, Matt Schwartz and Matt Strassler for discussions. This work is supported by the Swiss National Science Foundation (SNF) under the Sinergia grant number CRSII2 141847 1 and the US Department of Energy under grant DE-SC0013607. The authors thank the high-energy theory group at Harvard for hospitality and support.

\end{document}